%% file: ESCE-Letter-Resubmit_Final_Arxiv.tex
\documentclass[aps,prl,reprint,floatfix,showpacs]{revtex4-2}

\usepackage{amssymb}
\usepackage{amsmath}
\usepackage{amsfonts}
\usepackage{graphicx}

\newcommand{\fancy}{\mathcal}

\newcommand{\FE}{\kappa}

\renewcommand{\vec}[1]{\boldsymbol{#1}}

\newcommand{\Dp}[1]{\partial_{#1}}

\newcommand{\beq}{\begin{eqnarray}}
\newcommand{\eeq}{\end{eqnarray}}

\newcommand{\tr}{\text{Tr}}

\newcommand{\half}{\frac{1}{2}}

\newcommand{\rcite}[1]{Ref.~\onlinecite{#1}}

\newcommand{\Hx}{\text{Hx}}

\newcommand{\crm}{\text{c}}

\newcommand{\Hxc}{\text{Hxc}}

\newcommand{\EHxc}{{\cal E}_{\Hxc}}
\newcommand{\EHx}{{\cal E}_{\Hx}}

\newcommand{\Ec}{{\cal E}_{\crm}}

\newcommand{\Ts}{{\cal T}_{s}}
\newcommand{\T}{{\cal T}}
\newcommand{\E}{{\cal E}}
\newcommand{\F}{{\cal F}}

\newcommand{\Veeh}{\hat{V}_{\rm ee}}
\newcommand{\Vee}{{\cal V}_{\rm ee}}

\renewcommand{\vr}{\vec{r}}

\newcommand{\vR}{\vec{R}}

\newcommand{\wv}{\vec{w}}

\newcommand{\ext}{{\text{ext}}}

\newcommand{\ket}[1]{\left|#1\right\rangle}

\newcommand{\iket}[1]{|#1\rangle}

\newcommand{\ibraketop}[3]{\langle#1|#2|#3\rangle}
\newcommand{\ibkouter}[1]{|#1\rangle\langle#1|}

\newcommand{\iout}{\ibkouter}

\newcommand{\nh}{\hat{n}}
\newcommand{\Th}{\hat{T}}

\newcommand{\Hh}{\hat{H}}

\newcommand{\SCE}{\text{SCE}}
\newcommand{\ESCE}{V_{ee}^{\SCE}}

\newcommand{\ZPE}{\text{ZPE}}
\newcommand{\FZPE}{F^{\ZPE}}

\newcommand{\Gammah}{\hat{\Gamma}}

\usepackage[normalem]{ulem}
\usepackage{color}
\usepackage{xcolor}

\definecolor{Mygrey}{gray}{0.80}

\newcommand{\comment}[1]{}

\newcommand*{\LightComments}{}%

\newcommand{\RomanNumeralCaps}[1]
    {\MakeUppercase{\romannumeral #1}}
    
\definecolor{Mygrey}{gray}{0.80}
\definecolor{lteal}{rgb}{0.10,0.60,0.70}
\definecolor{dkred}{rgb}{0.80,0.10,0.00}
\definecolor{dkmagenta}{rgb}{0.70,0.00,0.70}

\newcommand\supp[1]{Supp. Mat. Sec.~\RomanNumeralCaps{#1}}

\ifdefined\NoComments
 \newcommand\TG[2]{#2}

\else
 \ifdefined\LightComments
  \newcommand\TG[2]{\textcolor{blue}{#2}}

 \else
  \newcommand\TG[2]{\textcolor{blue}{TG: \sout{#1}{#2}}}

 \fi
\fi

\begin{document}
\title{Electronic excited states in extreme limits via ensemble density functionals}

\author{Tim Gould}\affiliation{Queensland Micro- and Nanotechnology Centre, %
  Griffith University, Nathan, Qld 4111, Australia}
  \email{t.gould@griffith.edu.au}
\author{Derk P. Kooi}\affiliation{Department of Chemistry \& Pharmaceutical Sciences and Amsterdam Institute of Molecular and Life Sciences (AIMMS), Faculty of Science, Vrije Universiteit, De Boelelaan 1083, 1081HV Amsterdam, The Netherlands}
\author{Paola Gori-Giorgi}\affiliation{Department of Chemistry \& Pharmaceutical Sciences and Amsterdam Institute of Molecular and Life Sciences (AIMMS), Faculty of Science, Vrije Universiteit, De Boelelaan 1083, 1081HV Amsterdam, The Netherlands}
\author{Stefano Pittalis}\affiliation{CNR-Istituto Nanoscienze, Via
  Campi 213A, I-41125 Modena, Italy}

\begin{abstract}
Density functional theory (DFT) has greatly
expanded our ability to affordably compute and understand
electronic ground states, by replacing intractable {\em ab initio}
calculations by models based on paradigmatic physics
from high- and low-density limits.
But, a comparable treatment of excited states lags behind.
Here, we solve this outstanding problem by employing a generalization
of density functional theory to ensemble states (EDFT).
We thus address important paradigmatic cases of all
electronic systems in strongly (low-density) and weakly
(high-density) correlated regimes.
We show that the high-density limit connects to recent,
exactly-solvable EDFT results.
The low-density limit reveals an unnoticed and most unexpected result --
density functionals for strictly correlated {\em ground} states can be reused
{\em directly} for excited states. Non-trivial dependence
on excitation structure only shows up at third leading order.
Overall, our results provide foundations for effective models
of excited states that interpolate between exact
low- and high-density limits, which we illustrate on the
cases of singlet-singlet excitations in H$_2$ and
a ring of quantum wells.
\end{abstract}
\maketitle

{\em Preamble.~} 
Density functional theory (DFT) \cite{HohenbergKohn,KohnSham}
is best known as a computational
modelling tool used in tens of thousands of  {\em applicative} scientific papers every
year. What is less widely known is that DFT offers a natural connection between
quantum mechanics and paradigmatic physical conditions (high- and
low-density limits) of matter, in which electronic correlations
attain two quantitatively (weak and strong, respectively) and
qualitatively different fundamental ends.
In this context, DFT serves as a formal tool to understand
the behaviour of ground state electronic structure via a
rigorous constrained variational
approach to the electronic structure problem.
Understanding of paradigmatic conditions then
informs model development, e.g. the popular
``PBE''~\cite{DFA:pbepbe} approximation,
and computational studies therefrom.

Unfortunately, DFT is only defined for ground states,
so cannot elucidate the structure of excited states.
This Letter will demonstrate that {\em ensemble} density functional theory
(EDFT) for excited states~\cite{GOK-1,GOK-2}
can tackle this outstanding problem.
We shall show that recently derived Hartree and exchange physics%
~\cite{Gould2017-Limits,Gould2020-FDT}
become exact in the high density (weak interaction);
so high-density excited electronic states may be solved using these tools.
More importantly, we shall show that
the \emph{low density (strong interaction) limit of excited states
behaves exactly like a ground state}. Therefore,
the full suite of {\em ground state} strictly correlated electron
(SCE) tools and approximations%
~\cite{SeiGorSav-PRA-07,GorVigSei-JCTC-09,Lew-CRM-18,CotFriKlu-ARMA-18,FriGerGor-arxiv-22,VucGerDaaBahFriGor-arxiv-22}
may be used to solve both
ground and excited states of low-density many-electron systems. 

Our work thereby improves understanding of excited states in paradigmatic
limits and connects their behaviour to well-defined density functionals
for which exact forms and approximations are available.
It presents a crucial step toward efficient excited
state approximations that capture important limits;
and promises to accelerate and generalize
recent progress on low cost modelling of single%
~\cite{Filatov2015-Review,Gould2018-CT,Loos2020-EDFA,Gould2020-Molecules,Gould2021-EGKS,Gould2022-HL}
and double excitations%
~\cite{Filatov2015-Double,Sagredo2018,Loos2020-EDFA,Marut2020,Gould2021-DoubleX} 
that may range from weakly to strongly correlated regimes.

The rest of this Letter proceeds as follows: First, we introduce EDFT
and show how it can be used to understand the high-
and low-density limits of interacting electrons in realistic inhomogeneous systems. 
Then, using as an illustration the strong interaction limit
of electrons in an harmonic well, we derive the asymptotic properties
of the density functionals for describing excitations in Wigner-like
systems via EDFT. 
We then reveal that the \emph{second leading}
term in the low-density limit is also the same in ground and excited
states, and that a non-trivial dependence only appears
in the \emph{third leading} term -- which therefore
describe more realistic correlated excitations. 
We then illustrate the importance of our findings
for applications by studying excitations in two examples.
Finally, we conclude.

{\em Theoretical framework.~} Excited state
EDFT is concerned with the behaviour
of countable sets of excited states. In practice,
a finite set of low-lying solutions of $\hat{H}
\ket{\FE} = E_{\FE} \iket{\FE}$. These are grouped in an
ensemble state $\Gammah^{\wv}=\sum_{\FE}w_{\FE}\iout{\FE}$
using some prescribed weights~
\footnote{This use of prescribed weights
excludes the case of finite temperature
(thermal) ensembles,~\cite{Mermin1965}
where weights \emph{do depend} on energies
(and densities) in a non-trivial way}
such that $w_{\FE}\geq 0$ and
$\sum_{\FE} w_{\FE} = 1$ (collectively, $\wv$).
The average of an operator, $\hat{O}$,  over $\Gammah^{\wv}$ is
given by $\fancy{O}^{\wv}:=\tr[ \Gammah^{\wv} \hat{O} ]$. 
Crucially, choosing  $w_{\FE}\leq w_{\FE'}$ for
$E_{\FE}\geq E_{\FE'}$, ensures that $\Gammah^{\wv}$ fulfills an extended
variational principle~\cite{GOK-1} according to which
$\E^{\wv}=\inf_{\Gammah_{\rm trial}^{\wv}}\tr[\Gammah_{\rm trial}^{\wv}\Hh]$
where the argument for the infimum (usually a minimum),
$\Gammah^{\wv}_{\rm trial} = \sum_{\FE}w_{\FE}\iout{\FE_t}$,
involves prescribed weights, $\wv$, and mutually orthonormal trial
wavefunctions $\iket{\FE_t}$.

Density functionalizing the above variational principle in terms of
the ensemble particle density, $n$, yields,~\cite{GOK-2,Gould2017-Limits}
\begin{align}
  \E^{\wv} &= \min_{n}\bigg\{\T_s^{\wv}[n] + \EHxc^{\wv}[n]
  + \int n(\vr)v_{\ext}(\vr) d\vr\bigg\}.
  \label{eqn:EDFT}
\end{align}
Here,
$\T_s^{\wv}[n]=\min_{\Gammah^{\wv}_{\text{trial}}\to n}
\tr[\Gammah^{\wv}_{\text{trial}}\Th]$
is the kinetic energy of the Kohn-Sham (KS) system --
i.e., an auxiliary systems reproducing the
particle density of the ensemble; the minimum is attained at
$\Gammah_s\equiv \sum_{\FE}w_{\FE}\iout{\FE_s}$.%
~\footnote{We consider only ``well-behaved'' densities here
for which {$v_s[n]$} exists.}
$\EHxc^{\wv}[n]$ takes care of the remaining
Hartree (H), exchange (x), and correlation (c) energies.
Together, they yield the ``universal'' functional for ensembles:
$\F^{\wv}=\Ts^{\wv}+\EHxc^{\wv}$.

In fact, eq.~\eqref{eqn:EDFT} describes
{\em different functionals} for every choice of $\wv$.
To stress this important point we use capital calligraphic
letters to refer to energies of mixed states; and superscripts,
$\wv$, to indicate quantities that explicitly depend on their weights.
{\em Pure} ground states 
involve setting $w_0=1$ ($\Gammah^0=\iout{0}$) for which
eq.~\eqref{eqn:EDFT} attains usual DFT forms.
Spin and spatial symmetries are preserved at the
Kohn-Sham level~\cite{Gould2020-SP}
(see high-density limit discussion for further details)
by equally weighting degenerate states.
Varying weights (e.g. via partial derivatives)
lets us address {\em individual} excited states.~%
\footnote{Note, to address degenerate states
one must vary degenerate manifolds so that
degenerate states remain equally weighted.~\cite{Gould2020-SP}
E.g., addressing the first excited state of Be involves setting,
$\Gammah =\allowbreak (1-w)\iout{1s^22s^2}\allowbreak
+\tfrac{w}{3}(\iout{1s^22s2p_x}\allowbreak
+\iout{1s^22s2p_y}\allowbreak
+\iout{1s^22s2p_z})$,
and varying $w$. Then, perturbation theory is well-defined
around the ensemble Hx~\cite{Gould2017-Limits}
($\lambda\to 0^+$) limit.}
Therefore, the
weight-dependence of ensemble functionals is directly
related to the structure and behaviour of
ground and excited electronic states.

The universal energy functional may be generalized to,
\begin{align}
\F^{\lambda,\wv}[n] = \inf_{\Gammah^{\wv}\to n}\tr\big[\Gammah^{\wv}(\Th+\lambda\Veeh)\big]\;,
\label{eqn:F}
\end{align}
where $\Th$ is the kinetic energy operator and
$\Veeh$ is the Coulombic interaction operator.
$\lambda$ sets the strength of the interaction.
This functional is referred to as ``universal'' because the
external potential, which specifies the molecule or solid we wish to treat, does
not appear explicitly in its definition [the density being given
and fixed in eq.~\eqref{eqn:F}].
Matching terms from above yields $\F^{\wv}\equiv \F^{\lambda=1,\wv}$,
$\Ts^{\wv}\equiv \F^{0,\wv}$ and
$\EHxc^{\wv}:=\F^{\wv}-\F^{0,\wv}$.
In fact, we stress that $\F$, $\Ts$ and $\EHxc$ are \emph{multi}-universal because
each set of weights, $\wv$, defines a different excitation structure
and, thus, a different universal functional. As we
shall show below, this {\em multi}-universality
evolves into a {\em simple} universality in the low-density
limit of matter.

In what follows, our main objective is to determine the salient behavior
of key ensemble density functionals in the high- (weakly interacting)
and low-density (strictly interacting) limits. In doing so, we shall extend to excited states 
concepts and core results which have previously been worked out for pure ground states only.~\cite{SeiGorSav-PRA-07,GorVigSei-JCTC-09,Lew-CRM-18,CotFriKlu-ARMA-18,FriGerGor-arxiv-22,VucGerDaaBahFriGor-arxiv-22}
These works can be understood as providing a generalization of the seminal work of Wigner~\cite{Wig-PR-34,Wig-TFS-38} to {\em inhomogeneous} systems within DFT.
Our current work completes the generalization to include {\em excited} inhomogeneous
systems within EDFT. It thus provides a complete treatment
of electronic structure of two important paradigmatic and
fundamentally different regimes, within a consistent and
versatile approach.

{\em High-density limit.~}
In the parlance of modern density functional, the high- and low-density
limits entail uniform scaling of the coordinates of the electrons,
say, by $\gamma > 0$ in such a way  $n(\vr)\to \gamma^3 n(\gamma \vr)=:n_{\gamma}(\vr)$.
To keep the discussion simple, we may think of a finite system
like an atom, molecule or quantum dot. 
Scaling gives $\Th\to\gamma^{-2}\Th_{\gamma}$ and
$\Veeh\to \gamma^{-1}\hat{V}_{{\rm ee},\gamma}$; so,
\begin{align}
   \F^{\lambda,\wv}[n_\gamma] = \gamma^2 \F^{\lambda/\gamma,\wv}[n]\;.
   \label{eqn:FldUniv}
\end{align}
Because the scaled ensemble density is the density of a stationary ensemble 
of the Hamiltonian with interaction $\lambda = 1/\gamma$,
we see that the  high- and low-density limits are related to 
the weak- and strong-interaction limits, respectively.%
~\footnote{Note, we do not scale the mixing weights. Also note that,
level crossings in the ensemble are not a concern
as excitation energies do not change order under uniform scaling.}

\begin{figure}[t]
\includegraphics[width=\linewidth]{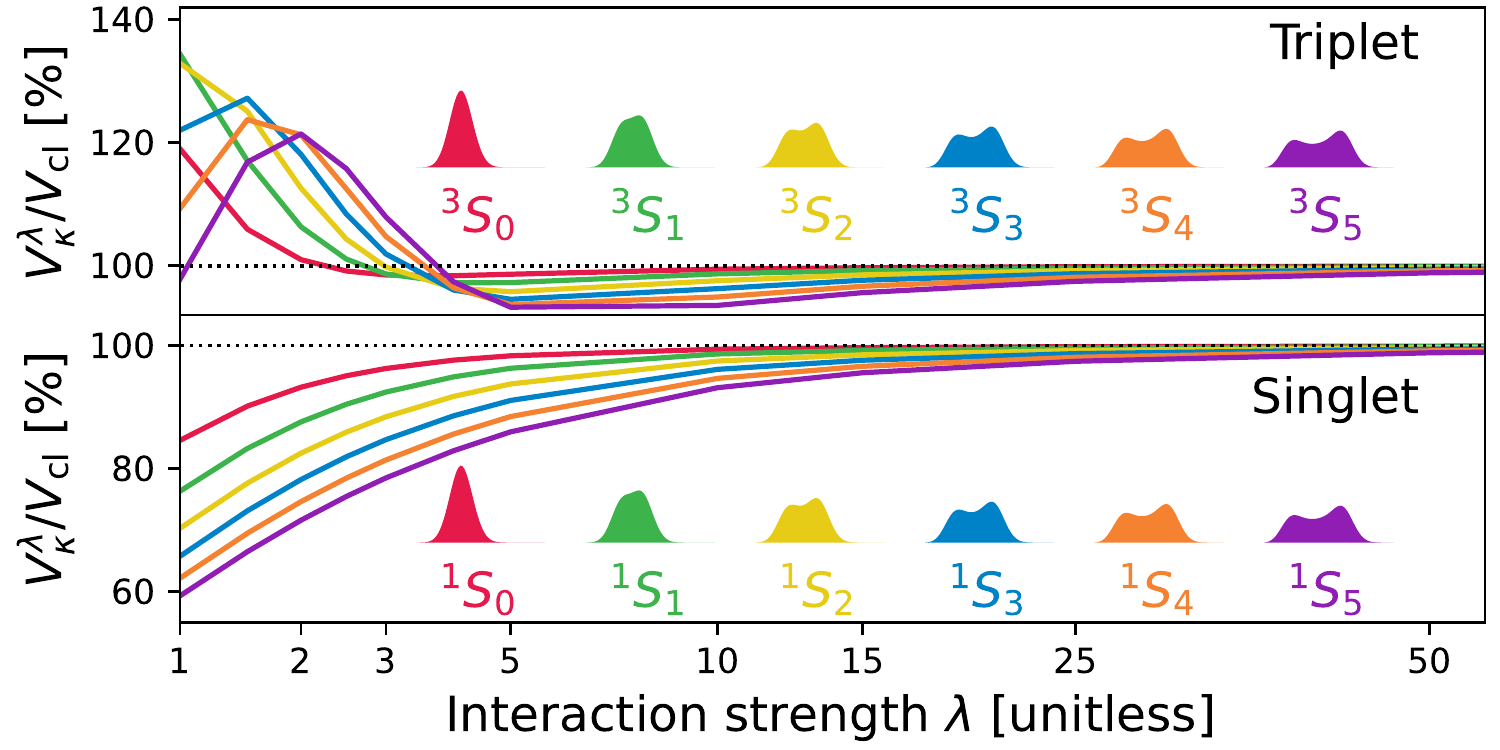}
\caption{Ratio of quantum and classical interaction energies for two
electrons in an Harmonic well. Shows six triplet (top) and
singlet (bottom) energies. Scaled densities, $4\pi r^2n(r)$,
of the states are also shown for the case $\lambda=50$.
\label{fig:Wint}}
\end{figure}

Let us first consider the high-density (i.e., weak interaction,
$\gamma\to\infty$)
limit. Scaling yields,
\begin{align}
\lim_{\gamma\to \infty} \tfrac{\F^{\wv}[n_\gamma]}{\gamma^2}  & =  \T^{\wv}_s[n]\;,
&
\lim_{\gamma\to \infty} \tfrac{\EHxc^{\wv}[n_{\gamma}]}{\gamma} & = \EHx^{\wv}[n]\;,
\label{eqn:hd}
\end{align}
where the second result follows from the definition~\cite{Gould2017-Limits}
of $\EHx$ as a gradient of $\F$.
The high-density limit thus
inherits good properties of $\EHx$:~%
\cite{Gould2017-Limits,Gould2020-FDT,Gould2020-SP,Gould2021-DoubleX}
i) it preserves spin and spatial symmetries of the system;
ii) the relevant KS states can be linear combinations
of Slater determinants (SD) that are  eigenstates of spin and
proper generators of point groups,
unlike the single SD of conventional
spin-DFT treatments; 
iii) yet, it enable effective reuse
of conventional spin-density functional approximations
for exchange, via combination rules or on-top pair densities.~%
\cite{Gould2020-FDT,Gould2020-Molecules,%
Gould2021-DoubleX,Gould2022-HL}

Next, consider the adiabatic connection formula,
\begin{align}
\EHxc^{\wv}[n]  =& \F^{\wv}[n]-\F^{0,\wv}[n]
=\int_0^1 \Vee^{\lambda,\wv}[n] d\lambda\;,
\label{eqn:ACF}
\\
\Vee^{\lambda,\wv}[n]=&\lim_{\eta\to 0^+}
\frac{\F^{\lambda+\eta,\wv}-\F^{\lambda,\wv}}{\eta}
:=\tr[\Veeh\Gammah^{\lambda^+,\wv}]\;.
\label{eqn:Vlambda}
\end{align} 
The `Hx' component is recovered as
$\EHx^{\wv}[n]  = \Vee^{0,\wv}[n]$.
Scaling gives
$\EHxc^{\wv}[n_{\gamma}]
=\gamma^2\int_0^{1/\gamma} \Vee^{\lambda,\wv}[n] d\lambda$,
from which (for finite systems) we get,~%
\footnote{This result is easily shown
by taking $\gamma^2$ times
a series expansion of eq.~\eqref{eqn:ACF} in small
$\lambda=\gamma^{-1}$.}
\begin{align}
\F^{\wv}[n_{\gamma}]
\underset{\gamma\to \infty}{\longrightarrow}
\gamma^2\Ts^{\wv}[n]+\gamma \EHx^{\wv}[n]
+\Ec^{\text{GL2},\wv}[n]+\ldots
\label{eqn:Fhdseries}
\end{align}
The first correlation contribution
follows from G\"orling-Levy~\cite{Goerling1993}
perturbation theory for ensembles,~\cite{Yang2021}
which must also be adapted for KS states in the form of linear combinations
of SDs.
Correlations may alternatively be captured by employing
expressions previously reported in
\rcite{Gould2019-DD,Gould2020-FDT,Fromager2020-DD};
as in (e.g.) \rcite{Gould2021-DoubleX}.

{\em Low-density limit.~}
Approaching a most striking, and previously unnoticed fact, let us
turn to the the low-density (i.e. strong interaction, $\gamma\to 0^+$)
limit,
\begin{align}
\lim_{\gamma \rightarrow 0^+} \F^{1, \wv}[n_\gamma]
 =  \lim_{\gamma \rightarrow 0^+}  \Vee^{1,\wv}[n_\gamma]\;.
\label{eqn:Fld}
\end{align}
Crucially,
\begin{align}
\lim_{\gamma \rightarrow 0^+} \tfrac{\Vee^{1,\wv}[n_\gamma]}{\gamma}
= \Vee^{\SCE ,\wv}[n] \equiv \ESCE[n]
\label{eqn:ESCE}
\end{align}
where $\ESCE[n]=\inf_{\Psi\to n}\ibraketop{\Psi}{\Veeh}{\Psi}$
is the known interaction energy functional of
strictly correlated electrons in a \emph{ground state}, but here
evaluated at the ensemble particle density.
This result says that, in the low-density limit, the functional
dependence on weights disappears from both
$\F^{1, \wv}[n_\gamma]$ and $\Vee^{1,\wv}[n_\gamma]$.
Dependence on the weights enters only via the particle density,
$n:=\tr[\Gammah^{\wv}\nh]=\sum_{\FE}w_{\FE}n_{\FE}$,
of the ensemble.
Eq.~\eqref{eqn:Fld},
and its extension to higher orders in $\gamma$
discussed later, are the central result of this work.
In this context, SCE results, analysis and understanding for ground
states~\cite{SeiGorSav-PRA-07,GorVigSei-JCTC-09,Lew-CRM-18,CotFriKlu-ARMA-18}
become special cases of the above more general result.

{\em Proof of Eqs~\eqref{eqn:Fld} and \eqref{eqn:ESCE}}.~
Here, we shall guide the reader through the main steps
and key physics. A full proof is reported in
Section~1 of the Supplementary Material (\supp{1}).

The salient features can be already
grasped by observing the behaviour of an interacting
system as interactions are increased in a model system.
We choose two-electron Harmonium in which two
electrons interact in an external potential $v_{\ext}=\half r^2$
with an interaction strength $\lambda$.
The scaled classical interaction energy of this system is
$V_{\rm cl}=0.7937 \lambda^{2/3}$.%
~\footnote{This is $V_{\rm cl}=\tfrac{\lambda}{2R_0}$ where $R_0$
minimizes the classical energy,
$E_{\rm cl}(R)=2\times\half R^2 + \tfrac{\lambda}{2R}$,
of two electrons interacting with
$\tfrac{\lambda}{|\vR_1-\vR_2|}$ when the
two electrons are on opposite sides of the well.}
Quantum solutions may be found numerically.
Details are in \supp{2}.
Figure~\ref{fig:Wint} shows the interaction energies,
$V_{\FE}^{\lambda}:=\ibraketop{\FE}{\Veeh}{\FE}$, of
six low-lying spherically symmetric triplet ($^3S$) and singlet ($^1S$)
states. It is clear that quantum and classical interaction energies
all become the same as $\lambda$ is increased -- i.e.,
all excitations tend toward the same classical limit.

\newcommand{\vs}{\vec{s}}
\newcommand{\vf}{\vec{f}}
To prove our result for EDFT we need to consider a similar
physical setting ($\lambda\to\infty$), in which instead
of fixing the external potential we fix the ensemble
density, containing the excited states we want to treat.
Proving eq.~\eqref{eqn:ESCE} then entails showing that the degeneracy behaviour carries through
to systems in which the density is kept fixed.
Our argument involves the expansion of wave functions
for large but finite interaction strengths,~\cite{GorVigSei-JCTC-09,ColDMaStra-arxiv-21}
around the strictly correlated limit. In this
effectively classical limit, which yields the leading term
as $\lambda\to\infty$ of the ground-state universal
functional $F^{\lambda}[n]$ \cite{Lew-CRM-18,CotFriKlu-ARMA-18},
the $N$-body distribution of an $N$-electron system is
$P_N[n](\vr_1\cdots\vr_N)
=\int \frac{n(\vs)}{N}\prod_{i=1}^N\delta(\vr_i-\vf_i(\vs))d\vs$
which leads to $F^{\lambda\to\infty}[n]\to\lambda \ESCE[n]
=\lambda \sum_{i=2}^N\int n(\vr) \tfrac{d\vr}{2|\vr-\vf_i(\vr)|}$.
Here, $\vf_i(\vr)$ are co-motion maps that preserve the density and the indistinguishability
of electrons.~\cite{SeiGorSav-PRA-07,FriGerGor-arxiv-22}

At large but finite $\lambda$ we
construct orthonormal wave functions,
$\iket{\FE^{\lambda}}$, based on
quantum harmonic oscillations (QHO)
around the strictly-correlated
distribution, $P_N[n]$.~\cite{GorVigSei-JCTC-09,Grossi2019}
The QHOs act on curvilinear coordinates
orthogonal to the manifold parametrized by the co-motion functions;
and contribute at $O(\sqrt{\lambda})$ in the
kinetic and potential energies. \emph{Prima facie}, the
wave functions $\iket{\FE^{\lambda}}$
have different densities. However,
it is also possible to quantize {\em along} the manifold,
which contributes only at $O(1)$ in kinetic energy
and is amenable to
the Harriman construction~\cite{Harriman1981} of orthogonal
orbitals yielding density $n$.
We thereby obtain a countable number of orthornomal
wavefunctions that all have the same density $n$, and
the same energy up to $O(1)$.
Thus, $\Vee^{\SCE,\wv}[n]\allowbreak
:=\sum_{\FE}w_{\FE}\ESCE[n]\allowbreak =\ESCE[n]$
and eqs~\eqref{eqn:Fld} and \eqref{eqn:ESCE} follow
from the equivalence of
$\gamma\to 0^+$ and $\lambda\to\infty$ in eq.~\eqref{eqn:FldUniv}.

{\em  Next-leading terms in the low-density limit.~}
To analyze the next leading terms, it useful to first rewrite
$\F^{\wv}[n]:=\T^{\SCE,\wv}[n] + \ESCE[n]$,
taking SCE as the reference system and letting
$\T^{\SCE,\wv}$ capture all the ensemble effects.
Then an alternative adiadatic connection yields,
\begin{align}
\T^{\SCE,\wv}[n]= \int_1^{\infty}  \frac{\mathcal{T}^{\lambda,\wv}[n]}{\lambda^2}d\lambda\;.
\label{eqn:TSCE}
\end{align}
Here, we introduced,
$\T^{\lambda,\wv}=-\lambda^2\Dp{\lambda}\tfrac{\F^{\lambda,\wv}}{\lambda}
:=\tr[\Th\Gammah^{\lambda^+,\wv}]$,
where the derivative and trace must be treated with caution,
like in eq.~\eqref{eqn:Vlambda}.
Eq.~\eqref{eqn:TSCE} generalizes known results for the
ground state-only case~\cite{Sav-PRA-95,LevGor-PRA-95,TeaHelSav-JCCS-16}
to the ensembles considered in this work.

Next, we show that $\T^{\lambda\to\infty,\wv}$
is independent of weights to leading order,
which leads to $\T^{\SCE,\wv}[n_{\gamma\to 0^+}]$ also independent of weights.
Ensemblization of known results~\cite{GorVigSei-JCTC-09,GorSei-PCCP-10,Grossi2019} gives
$\T^{\lambda\to\infty,\wv}\to\tfrac{\sqrt{\lambda}}{2}\F^{\ZPE,\wv}$
where $\F^{\ZPE,\wv}$ involves $\lambda$-normalized
zero point energy (ZPE)
of the QHOs, $\iket{\FE^{\lambda}}$, introduced earlier.
Hence, $\T^{\lambda,\wv}$ becomes independent of weights
if we can show that $\F^{\ZPE,\wv}[n] \equiv \lim_{\lambda\to\infty}
\tfrac{2}{\sqrt{\lambda}}\T^{\lambda,\wv}[n]$
is independent of weights.
\supp{1}
naturally covers this case -- for guidance, below, we touch on
essential steps and consequences.

The Harriman construction introduced earlier 
(also, \supp{1B}) yields
$\ibraketop{\FE^{\lambda}}{\Th}{\FE^{\lambda}}
=\ibraketop{0^{\lambda}}{\Th}{0^{\lambda}}+O(1)$
for $\lambda\to\infty$. Thus, 
$\T^{\lambda,\wv}=\sum_{\FE}w_{\FE}\ibraketop{\FE^{\lambda}}{\Th}{\FE^{\lambda}}=\ibraketop{0^{\lambda}}{\Th}{0^{\lambda}}+O(1)$
is independent of weights to leading order, giving
weight-independent,
$\F^{\ZPE,\wv}[n]
\equiv \lim_{\lambda\to\infty}\tfrac{2\T^{\lambda,\wv}[n]}{\sqrt{\lambda}}
=\FZPE[n]$.
Here, $\FZPE[n]$ is the well-studied ground state functional%
~\cite{GorVigSei-JCTC-09,Grossi2019,ColDMaStra-arxiv-21},
but evaluated on the ensemble density.
Using the latter result in \eqref{eqn:TSCE}, and
applying scaling laws, finally yields,
$\lim_{\gamma\to 0^+}\T^{\SCE,\wv}[n_{\gamma}]=
\FZPE[n_{\gamma}] =\gamma^{3/2}\FZPE[n]$.
Thus we conclude that $\T^{\lambda,\wv}[n]$ and $\T^{\SCE,\wv}[n]$ are
independent of the ensemble weights in
the low-density limit. 
$\F^{\wv}[n_{\gamma\to 0^+}]=\gamma\ESCE[n]+
\gamma^{3/2}\FZPE[n]$ is therefore
also independent of weights to second leading order.
Details of scaling are in \supp{3}. 

Where and how does the
weight dependence appears in the low-density limit?
Eqs~(S31)--(S37) of \supp{1B} reveal that it
appears in the third leading term,
\begin{align}
\F^{\wv}[n_{\gamma}]
\underset{\gamma\to 0^+}{\longrightarrow}&
\gamma\ESCE[n]+\gamma^{3/2}\FZPE[n]
\nonumber\\&
+ \gamma^2\Delta\T^{(2),\wv}[n] + \ldots \;.
\label{eqn:Fldseries}
\end{align}
The $O(\gamma^2)$ term,
$\Delta\T^{(2),\wv}=\sum_{\FE}w_{\FE}\Delta T^{(2)}_{\FE}$,
has an explicit weight dependence on each excited state.
It captures the energy of oscillations ``perpendicular''
to the collective ZPE modes, according to the metric dictated by
the SCE manifold -- see eq.~(S36) and discussion for details.
Note, a similar result was previously observed in
the special case of Hubbard dimers.~\cite{Deur2018}

\begin{figure}[h!t]
\includegraphics[width=\linewidth]{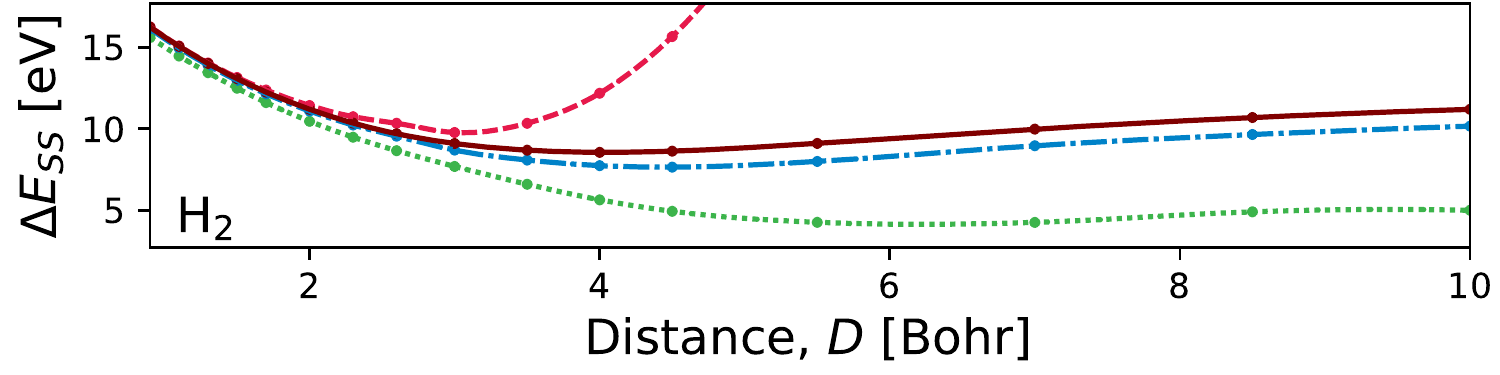}\\
\includegraphics[width=\linewidth]{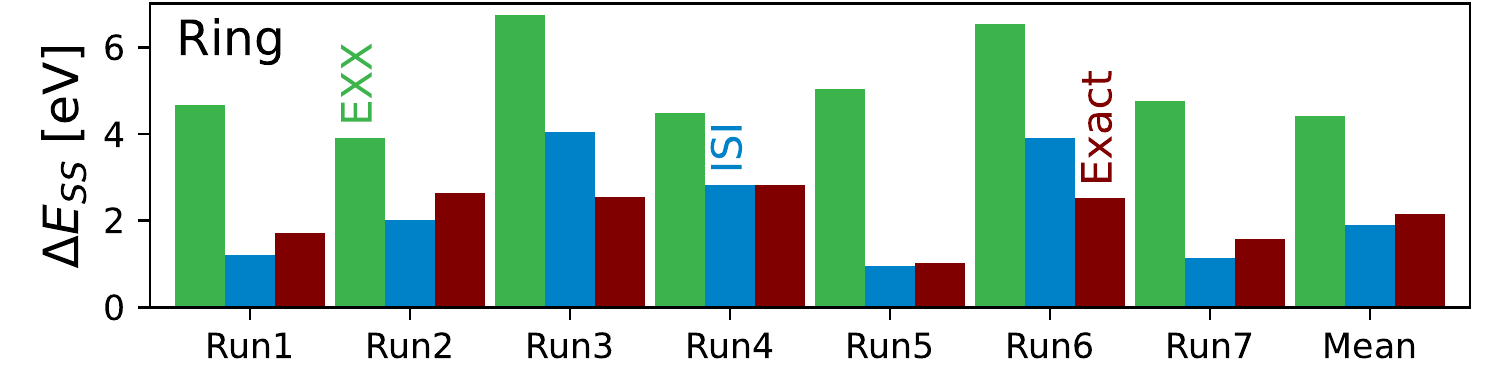}\\
\caption{
Excitation energies, for dissociating H$_2$ (top); and
seven random runs and the mean of 25 runs
for a ring of wells with disorder (bottom).
Shows ensemblized EXX (dots/green), 
GL2 (dashes/red), ISI (dash-dot/blue) and
exact (lines/maroon) energies. 
GL2 errors in the ring are so large they cannot be
shown in the lower panel.
\label{fig:Models}}
\end{figure}

{\em Use of high- and low-density limits in approximations.~}
We have so far derived series expansions in the high-density
[eq.~\eqref{eqn:Fhdseries}] and low-density
[eq.~\eqref{eqn:Fldseries}] limits.
Next, we  shall illustrate their relevance in applications.

First, we remark that eqs~\eqref{eqn:ACF}, \eqref{eqn:Fhdseries}
and \eqref{eqn:Fldseries} imply that,
\begin{align}
\lim_{\gamma\to +\infty} \EHxc^{\wv}[n_{\gamma}] & =  
\gamma \EHx^{\wv}[n] +\Ec^{\text{GL2},\wv}[n]+\ldots\;,
\label{eqn:EHxchd}
\\
\lim_{\gamma\to 0^+} \EHxc^{\wv}[n_{\gamma}] & =  
\gamma \ESCE[n] + \gamma^{3/2}F^{\ZPE}[n] +\ldots \;.
\label{eqn:EHxcld}
\end{align}
Especially note that the low-density (strictly correlated)
limit of both $\F^{\wv}[n]$ [see \eqref{eqn:Fld}]
and $\EHxc^{\wv}[n]$
depend on the excitation structure only trivially,
via the ensemble particle density.
Weight dependence appears at higher order.

To illustrate the usefulness of the
limits, we first consider the lowest
singlet-singlet excitation in dissociating
H$_2$. This problem is a rather stringent test of 
density functionals -- failed by time-dependent
DFT approximations~\cite{Maitra2022-AR,Giesbertz2008-CTDBB}
-- because:
i) the ground state is dynamically correlated
near its minima but becomes strongly correlated when
dissociated;
ii) the first excited state is always dynamically
correlated, thus cancellation of errors
in the approximate excitation energy from the
ground state may be unreliable during dissociation;
iii) the first excited state in the KS ensemble
involves a superposition of two SDs, and its
symmetry and related properties are irreproducible
by an adiabatic single-SD approach.

Figure~\ref{fig:Models} (top) reports 
the excitation energy $\Delta E_{SS}=E_{S_1}-E_{S_0}$
of the lowest singlet states ($S_0$ and $S_1$) of
H$_2$ using:
EXact eXchange (EXX) energies -- the
leading term in the \emph{high}-density series
of eq.~\eqref{eqn:EHxchd};
G\"orling-Levy~\cite{Goerling1993}
(GL2) perturbation theory -- the next
leading term of \eqref{eqn:EHxchd}; and
the Interaction Strength Interpolation (ISI)
approximation~\cite{Seidl2000,SDGF22}
that uses high- and low-density limits --
the latter via the harmonium point charge plus
continuum (hPC) approximation~\cite{SDGF22}.
Note, all approximations are {\em ensemblized}
versions of ground states analogs,
i.e. $E_{\Hxc}^{\text{approx}}\to\EHxc^{\text{approx},\wv}$
is adapted for excited states.
All relevant energy expressions and technical details
on the calculations are in \supp{4}.

Only ISI performs well across the whole H$_2$
dissociation curve, which unambiguously highlights
the benefit of using
both eq.~\eqref{eqn:EHxchd} and
eq.~\eqref{eqn:EHxcld}
to construct approximations that capture
different correlation regimes.
In fact, using only eq.~\eqref{eqn:EHxchd}
leads to very poor results for the ground state energy:
EXX overestimates and GL2 drastically
underestimates as correlations become stronger.

Next, we carry out similar calculations for
four electrons in a ring of four quantum wells;
see \supp{4} for further details.
Lattice disorder in this system yields
$\Delta E_{SS}=2.15$~eV on average,
versus $\Delta E_{SS}=0.003$~eV
of the ordered lattice.
Results are shown in Figure~\ref{fig:Models} (bottom).
Again, we see that the low-density behaviour
included in ISI reduces errors: from
100\% (EXX) down to -12\% (ISI).
GL2 energies (not shown) have
orders of magnitude worse errors.
This example (also, \rcite{Marie2022})
thus suggests that seamless
interpolation between high- and low-density limits
may be crucial for predicting optical gaps
in disordered nanostructures.

{\em Summary and conclusions.~}
The results presented in this work describe,
via ensemble density functionals, the behaviour
of excited many-electron states in the paradigmatic
high- (weakly-) and low-density (strictly-correlated) limits (regimes)
-- summarized for the important $\EHxc$ functional
in eqs~\eqref{eqn:EHxchd} and \eqref{eqn:EHxcld},
respectively.

The high-density limit follows intuition and
connects directly to previous results which use the 
\emph{ensemble} Kohn-Sham system as reference system.
The corresponding auxiliary pure states have the form
of symmetry adapted combination of Slater
determinants; and $\EHxc$ has strong weight-dependence.
Approximations based on this limit have already successfully
described weakly to moderately correlated excitations.

The low-density limit, in contrast, revealed
an unexpected fact: the first two leading order terms
of excited states may be described by existing tools used
for strictly correlated {\em ground} states.
Therefore the ensemblization of ground-state approximations is,
for once, straightforward. Dependence of \TG{the (multi-)universal
functional forms}{$\EHxc$} on the weights (and therefore excitation
structure) only shows up in the third leading order term.
The provided model applications illustrate that
generally correlated regimes of excited
states requires seamless treatment of both density regimes.

One immediate consequence of the present
work is that electronic interaction models must interpolate
between Fermionic mean-field like excitation-structure dependence at
high-densities, and no excitation-structure dependence at low-densities.
Not only is this of direct importance for traditional
analytic-driven approximations, as seen in
the examples reported here, it also provides
constraints for data-driven methodologies based on machine learning.
Ensemble-derived constraints were used to great
success in the machine-learned ``Deep Mind 21''
ground state approximation~\cite{DM21}
-- our work promises to extend this success
to excited states.

Natural next steps from the present results
are to consider finite-temperature
ensembles and magnetic interactions.
Finite temperature imposes a
$\lambda$-dependence on the weights. Prior
work~\cite{Mermin1965, Pittalis2011-FT} showed that the
high/low-density limit may be more relevant to the behavior of
density functionals at low/high temperatures.
Magnetic interactions require extra
basic densities (e.g. spin-densities and currents)
and related Hxc quantities;~%
\cite{Vignale1988, Bencheikh_2003}
and must consistently fulfill gauge
symmetries.~\cite{Pittalis-U1SU2}
Further work along both lines is being pursued.

\acknowledgments

TG was supported by an Australian Research Council (ARC)
Discovery Project (DP200100033) and Future Fellowship (FT210100663).
DPK and PG-G were supported by the Netherlands Organisation for Scientific Research (NWO) under Vici grant 724.017.001.
SP was partially supported by the MIUR PRIN Grant No. 2017RKWTMY.

\input{ESCE-Letter-Resubmit_Final_Arxiv.bbl}

\end{document}

%% file: ESCE-Letter-Resubmit_Final_Arxiv.bbl
%

%% file: ESCE-Letter-Resubmit_Final_Arxiv.bbl
\begin{thebibliography}{55}%
\makeatletter
\providecommand \@ifxundefined [1]{%
 \@ifx{#1\undefined}
}%
\providecommand \@ifnum [1]{%
 \ifnum #1\expandafter \@firstoftwo
 \else \expandafter \@secondoftwo
 \fi
}%
\providecommand \@ifx [1]{%
 \ifx #1\expandafter \@firstoftwo
 \else \expandafter \@secondoftwo
 \fi
}%
\providecommand \natexlab [1]{#1}%
\providecommand \enquote  [1]{``#1''}%
\providecommand \bibnamefont  [1]{#1}%
\providecommand \bibfnamefont [1]{#1}%
\providecommand \citenamefont [1]{#1}%
\providecommand \href@noop [0]{\@secondoftwo}%
\providecommand \href [0]{\begingroup \@sanitize@url \@href}%
\providecommand \@href[1]{\@@startlink{#1}\@@href}%
\providecommand \@@href[1]{\endgroup#1\@@endlink}%
\providecommand \@sanitize@url [0]{\catcode `\\12\catcode `\$12\catcode
  `\&12\catcode `\#12\catcode `\^12\catcode `\_12\catcode `\%12\relax}%
\providecommand \@@startlink[1]{}%
\providecommand \@@endlink[0]{}%
\providecommand \url  [0]{\begingroup\@sanitize@url \@url }%
\providecommand \@url [1]{\endgroup\@href {#1}{\urlprefix }}%
\providecommand \urlprefix  [0]{URL }%
\providecommand \Eprint [0]{\href }%
\providecommand \doibase [0]{https://doi.org/}%
\providecommand \selectlanguage [0]{\@gobble}%
\providecommand \bibinfo  [0]{\@secondoftwo}%
\providecommand \bibfield  [0]{\@secondoftwo}%
\providecommand \translation [1]{[#1]}%
\providecommand \BibitemOpen [0]{}%
\providecommand \bibitemStop [0]{}%
\providecommand \bibitemNoStop [0]{.\EOS\space}%
\providecommand \EOS [0]{\spacefactor3000\relax}%
\providecommand \BibitemShut  [1]{\csname bibitem#1\endcsname}%
\let\auto@bib@innerbib\@empty
\bibitem [{\citenamefont {Hohenberg}\ and\ \citenamefont
  {Kohn}(1964)}]{HohenbergKohn}%
  \BibitemOpen
  \bibfield  {author} {\bibinfo {author} {\bibfnamefont {P.}~\bibnamefont
  {Hohenberg}}\ and\ \bibinfo {author} {\bibfnamefont {W.}~\bibnamefont
  {Kohn}},\ }\bibfield  {title} {\bibinfo {title} {Inhomogeneous electron
  gas},\ }\href@noop {} {\bibfield  {journal} {\bibinfo  {journal} {Phys Rev}\
  }\textbf {\bibinfo {volume} {136}},\ \bibinfo {pages} {B864} (\bibinfo {year}
  {1964})}\BibitemShut {NoStop}%
\bibitem [{\citenamefont {Kohn}\ and\ \citenamefont {Sham}(1965)}]{KohnSham}%
  \BibitemOpen
  \bibfield  {author} {\bibinfo {author} {\bibfnamefont {W.}~\bibnamefont
  {Kohn}}\ and\ \bibinfo {author} {\bibfnamefont {L.~J.}\ \bibnamefont
  {Sham}},\ }\bibfield  {title} {\bibinfo {title} {Self-consistent equations
  including exchange and correlation effects},\ }\href@noop {} {\bibfield
  {journal} {\bibinfo  {journal} {Phys Rev}\ }\textbf {\bibinfo {volume}
  {140}},\ \bibinfo {pages} {A1133} (\bibinfo {year} {1965})}\BibitemShut
  {NoStop}%
\bibitem [{\citenamefont {Perdew}\ \emph {et~al.}(1996)\citenamefont {Perdew},
  \citenamefont {Burke},\ and\ \citenamefont {Ernzerhof}}]{DFA:pbepbe}%
  \BibitemOpen
  \bibfield  {author} {\bibinfo {author} {\bibfnamefont {J.~P.}\ \bibnamefont
  {Perdew}}, \bibinfo {author} {\bibfnamefont {K.}~\bibnamefont {Burke}},\ and\
  \bibinfo {author} {\bibfnamefont {M.}~\bibnamefont {Ernzerhof}},\ }\bibfield
  {title} {\bibinfo {title} {Generalized gradient approximation made simple},\
  }\href@noop {} {\bibfield  {journal} {\bibinfo  {journal} {Phys Rev Lett}\
  }\textbf {\bibinfo {volume} {77}},\ \bibinfo {pages} {3865} (\bibinfo {year}
  {1996})}\BibitemShut {NoStop}%
\bibitem [{\citenamefont {Gross}\ \emph
  {et~al.}(1988{\natexlab{a}})\citenamefont {Gross}, \citenamefont {Oliveira},\
  and\ \citenamefont {Kohn}}]{GOK-1}%
  \BibitemOpen
  \bibfield  {author} {\bibinfo {author} {\bibfnamefont {E.~K.~U.}\
  \bibnamefont {Gross}}, \bibinfo {author} {\bibfnamefont {L.~N.}\ \bibnamefont
  {Oliveira}},\ and\ \bibinfo {author} {\bibfnamefont {W.}~\bibnamefont
  {Kohn}},\ }\bibfield  {title} {\bibinfo {title} {{Rayleigh-Ritz} variational
  principle for ensembles of fractionally occupied states},\ }\href@noop {}
  {\bibfield  {journal} {\bibinfo  {journal} {Phys Rev A}\ }\textbf {\bibinfo
  {volume} {37}},\ \bibinfo {pages} {2805} (\bibinfo {year}
  {1988}{\natexlab{a}})}\BibitemShut {NoStop}%
\bibitem [{\citenamefont {Gross}\ \emph
  {et~al.}(1988{\natexlab{b}})\citenamefont {Gross}, \citenamefont {Oliveira},\
  and\ \citenamefont {Kohn}}]{GOK-2}%
  \BibitemOpen
  \bibfield  {author} {\bibinfo {author} {\bibfnamefont {E.~K.~U.}\
  \bibnamefont {Gross}}, \bibinfo {author} {\bibfnamefont {L.~N.}\ \bibnamefont
  {Oliveira}},\ and\ \bibinfo {author} {\bibfnamefont {W.}~\bibnamefont
  {Kohn}},\ }\bibfield  {title} {\bibinfo {title} {Density-functional theory
  for ensembles of fractionally occupied states. i. basic formalism},\
  }\href@noop {} {\bibfield  {journal} {\bibinfo  {journal} {Phys Rev A}\
  }\textbf {\bibinfo {volume} {37}},\ \bibinfo {pages} {2809} (\bibinfo {year}
  {1988}{\natexlab{b}})}\BibitemShut {NoStop}%
\bibitem [{\citenamefont {Gould}\ and\ \citenamefont
  {Pittalis}(2017)}]{Gould2017-Limits}%
  \BibitemOpen
  \bibfield  {author} {\bibinfo {author} {\bibfnamefont {T.}~\bibnamefont
  {Gould}}\ and\ \bibinfo {author} {\bibfnamefont {S.}~\bibnamefont
  {Pittalis}},\ }\bibfield  {title} {\bibinfo {title} {{Hartree} and exchange
  in ensemble density functional theory: Avoiding the nonuniqueness disaster},\
  }\href@noop {} {\bibfield  {journal} {\bibinfo  {journal} {Phys Rev Lett}\
  }\textbf {\bibinfo {volume} {119}},\ \bibinfo {pages} {243001} (\bibinfo
  {year} {2017})}\BibitemShut {NoStop}%
\bibitem [{\citenamefont {Gould}\ \emph {et~al.}(2020)\citenamefont {Gould},
  \citenamefont {Stefanucci},\ and\ \citenamefont {Pittalis}}]{Gould2020-FDT}%
  \BibitemOpen
  \bibfield  {author} {\bibinfo {author} {\bibfnamefont {T.}~\bibnamefont
  {Gould}}, \bibinfo {author} {\bibfnamefont {G.}~\bibnamefont {Stefanucci}},\
  and\ \bibinfo {author} {\bibfnamefont {S.}~\bibnamefont {Pittalis}},\
  }\bibfield  {title} {\bibinfo {title} {Ensemble density functional theory:
  Insight from the fluctuation-dissipation theorem},\ }\href@noop {} {\bibfield
   {journal} {\bibinfo  {journal} {Phys Rev Lett}\ }\textbf {\bibinfo {volume}
  {125}},\ \bibinfo {pages} {233001} (\bibinfo {year} {2020})}\BibitemShut
  {NoStop}%
\bibitem [{\citenamefont {Seidl}\ \emph {et~al.}(2007)\citenamefont {Seidl},
  \citenamefont {Gori-Giorgi},\ and\ \citenamefont {Savin}}]{SeiGorSav-PRA-07}%
  \BibitemOpen
  \bibfield  {author} {\bibinfo {author} {\bibfnamefont {M.}~\bibnamefont
  {Seidl}}, \bibinfo {author} {\bibfnamefont {P.}~\bibnamefont {Gori-Giorgi}},\
  and\ \bibinfo {author} {\bibfnamefont {A.}~\bibnamefont {Savin}},\ }\bibfield
   {title} {\bibinfo {title} {Strictly correlated electrons in
  density-functional theory: A general formulation with applications to
  spherical densities},\ }\href {https://doi.org/10.1103/PhysRevA.75.042511}
  {\bibfield  {journal} {\bibinfo  {journal} {Phys Rev A}\ }\textbf {\bibinfo
  {volume} {75}},\ \bibinfo {pages} {042511/12} (\bibinfo {year}
  {2007})}\BibitemShut {NoStop}%
\bibitem [{\citenamefont {Gori-Giorgi}\ \emph {et~al.}(2009)\citenamefont
  {Gori-Giorgi}, \citenamefont {Vignale},\ and\ \citenamefont
  {Seidl}}]{GorVigSei-JCTC-09}%
  \BibitemOpen
  \bibfield  {author} {\bibinfo {author} {\bibfnamefont {P.}~\bibnamefont
  {Gori-Giorgi}}, \bibinfo {author} {\bibfnamefont {G.}~\bibnamefont
  {Vignale}},\ and\ \bibinfo {author} {\bibfnamefont {M.}~\bibnamefont
  {Seidl}},\ }\bibfield  {title} {\bibinfo {title} {Electronic zero-point
  oscillations in the strong-interaction limit of density functional theory},\
  }\href {https://doi.org/10.1021/ct8005248} {\bibfield  {journal} {\bibinfo
  {journal} {J Chem Theory Comput}\ }\textbf {\bibinfo {volume} {5}},\ \bibinfo
  {pages} {743} (\bibinfo {year} {2009})}\BibitemShut {NoStop}%
\bibitem [{\citenamefont {Lewin}(2018)}]{Lew-CRM-18}%
  \BibitemOpen
  \bibfield  {author} {\bibinfo {author} {\bibfnamefont {M.}~\bibnamefont
  {Lewin}},\ }\bibfield  {title} {\bibinfo {title} {Semi-classical limit of the
  {L}evy--{L}ieb functional in {D}ensity {F}unctional {T}heory},\ }\href
  {https://doi.org/10.1016/j.crma.2018.03.002} {\bibfield  {journal} {\bibinfo
  {journal} {C R Math}\ }\textbf {\bibinfo {volume} {356}},\ \bibinfo {pages}
  {449} (\bibinfo {year} {2018})}\BibitemShut {NoStop}%
\bibitem [{\citenamefont {Cotar}\ \emph {et~al.}(2018)\citenamefont {Cotar},
  \citenamefont {Friesecke},\ and\ \citenamefont
  {Kl{\"u}ppelberg}}]{CotFriKlu-ARMA-18}%
  \BibitemOpen
  \bibfield  {author} {\bibinfo {author} {\bibfnamefont {C.}~\bibnamefont
  {Cotar}}, \bibinfo {author} {\bibfnamefont {G.}~\bibnamefont {Friesecke}},\
  and\ \bibinfo {author} {\bibfnamefont {C.}~\bibnamefont {Kl{\"u}ppelberg}},\
  }\bibfield  {title} {\bibinfo {title} {Smoothing of transport plans with
  fixed marginals and rigorous semiclassical limit of the hohenberg--kohn
  functional},\ }\href {https://doi.org/10.1007/s00205-017-1208-y} {\bibfield
  {journal} {\bibinfo  {journal} {Arch. Ration. Mech. An.}\ }\textbf {\bibinfo
  {volume} {228}},\ \bibinfo {pages} {891} (\bibinfo {year}
  {2018})}\BibitemShut {NoStop}%
\bibitem [{\citenamefont {Friesecke}\ \emph {et~al.}(2022)\citenamefont
  {Friesecke}, \citenamefont {Gerolin},\ and\ \citenamefont
  {Gori-Giorgi}}]{FriGerGor-arxiv-22}%
  \BibitemOpen
  \bibfield  {author} {\bibinfo {author} {\bibfnamefont {G.}~\bibnamefont
  {Friesecke}}, \bibinfo {author} {\bibfnamefont {A.}~\bibnamefont {Gerolin}},\
  and\ \bibinfo {author} {\bibfnamefont {P.}~\bibnamefont {Gori-Giorgi}},\
  }\bibfield  {title} {\bibinfo {title} {The strong-interaction limit of
  density functional theory},\ }\href@noop {} {\bibfield  {journal} {\bibinfo
  {journal} {arXiv preprint arXiv:2202.09760}\ } (\bibinfo {year}
  {2022})}\BibitemShut {NoStop}%
\bibitem [{\citenamefont {Vuckovic}\ \emph {et~al.}(2022)\citenamefont
  {Vuckovic}, \citenamefont {Gerolin}, \citenamefont {Daas}, \citenamefont
  {Bahmann}, \citenamefont {Friesecke},\ and\ \citenamefont
  {Gori-Giorgi}}]{VucGerDaaBahFriGor-arxiv-22}%
  \BibitemOpen
  \bibfield  {author} {\bibinfo {author} {\bibfnamefont {S.}~\bibnamefont
  {Vuckovic}}, \bibinfo {author} {\bibfnamefont {A.}~\bibnamefont {Gerolin}},
  \bibinfo {author} {\bibfnamefont {T.~J.}\ \bibnamefont {Daas}}, \bibinfo
  {author} {\bibfnamefont {H.}~\bibnamefont {Bahmann}}, \bibinfo {author}
  {\bibfnamefont {G.}~\bibnamefont {Friesecke}},\ and\ \bibinfo {author}
  {\bibfnamefont {P.}~\bibnamefont {Gori-Giorgi}},\ }\bibfield  {title}
  {\bibinfo {title} {Density functionals based on the mathematical structure of
  the strong-interaction limit of dft},\ }\href@noop {} {\bibfield  {journal}
  {\bibinfo  {journal} {arXiv preprint arXiv:2204.10769}\ } (\bibinfo {year}
  {2022})}\BibitemShut {NoStop}%
\bibitem [{\citenamefont {Filatov}(2015)}]{Filatov2015-Review}%
  \BibitemOpen
  \bibfield  {author} {\bibinfo {author} {\bibfnamefont {M.}~\bibnamefont
  {Filatov}},\ }\bibfield  {title} {\bibinfo {title} {Spin-restricted
  ensemble-referenced {Kohn-Sham} method: Basic principles and application to
  strongly correlated ground and excited states of molecules},\ }\href@noop {}
  {\bibfield  {journal} {\bibinfo  {journal} {WIREs Comput Mol Sci}\ }\textbf
  {\bibinfo {volume} {5}},\ \bibinfo {pages} {146} (\bibinfo {year}
  {2015})}\BibitemShut {NoStop}%
\bibitem [{\citenamefont {Gould}\ \emph {et~al.}(2018)\citenamefont {Gould},
  \citenamefont {Kronik},\ and\ \citenamefont {Pittalis}}]{Gould2018-CT}%
  \BibitemOpen
  \bibfield  {author} {\bibinfo {author} {\bibfnamefont {T.}~\bibnamefont
  {Gould}}, \bibinfo {author} {\bibfnamefont {L.}~\bibnamefont {Kronik}},\ and\
  \bibinfo {author} {\bibfnamefont {S.}~\bibnamefont {Pittalis}},\ }\bibfield
  {title} {\bibinfo {title} {Charge transfer excitations from exact and
  approximate ensemble {Kohn-Sham} theory},\ }\href@noop {} {\bibfield
  {journal} {\bibinfo  {journal} {J Chem Phys}\ }\textbf {\bibinfo {volume}
  {148}},\ \bibinfo {pages} {174101} (\bibinfo {year} {2018})}\BibitemShut
  {NoStop}%
\bibitem [{\citenamefont {Loos}\ and\ \citenamefont
  {Fromager}(2020)}]{Loos2020-EDFA}%
  \BibitemOpen
  \bibfield  {author} {\bibinfo {author} {\bibfnamefont {P.-F.}\ \bibnamefont
  {Loos}}\ and\ \bibinfo {author} {\bibfnamefont {E.}~\bibnamefont
  {Fromager}},\ }\bibfield  {title} {\bibinfo {title} {A weight-dependent local
  correlation density-functional approximation for ensembles},\ }\href@noop {}
  {\bibfield  {journal} {\bibinfo  {journal} {J. Chem. Phys,}\ }\textbf
  {\bibinfo {volume} {152}},\ \bibinfo {pages} {214101} (\bibinfo {year}
  {2020})}\BibitemShut {NoStop}%
\bibitem [{\citenamefont {Gould}(2020)}]{Gould2020-Molecules}%
  \BibitemOpen
  \bibfield  {author} {\bibinfo {author} {\bibfnamefont {T.}~\bibnamefont
  {Gould}},\ }\bibfield  {title} {\bibinfo {title} {Approximately
  self-consistent ensemble density functional theory: Toward inclusion of all
  correlations},\ }\href@noop {} {\bibfield  {journal} {\bibinfo  {journal} {J
  Phys Chem Lett}\ }\textbf {\bibinfo {volume} {11}},\ \bibinfo {pages} {9907}
  (\bibinfo {year} {2020})}\BibitemShut {NoStop}%
\bibitem [{\citenamefont {Gould}\ and\ \citenamefont
  {Kronik}(2021)}]{Gould2021-EGKS}%
  \BibitemOpen
  \bibfield  {author} {\bibinfo {author} {\bibfnamefont {T.}~\bibnamefont
  {Gould}}\ and\ \bibinfo {author} {\bibfnamefont {L.}~\bibnamefont {Kronik}},\
  }\bibfield  {title} {\bibinfo {title} {Ensemble generalized {Kohn}-{Sham}
  theory: The good, the bad, and the ugly},\ }\href
  {https://doi.org/10.1063/5.0040447} {\bibfield  {journal} {\bibinfo
  {journal} {J Chem Phys}\ }\textbf {\bibinfo {volume} {154}},\ \bibinfo
  {pages} {094125} (\bibinfo {year} {2021})}\BibitemShut {NoStop}%
\bibitem [{\citenamefont {Gould}\ \emph {et~al.}(2022)\citenamefont {Gould},
  \citenamefont {Hashimi}, \citenamefont {Kronik},\ and\ \citenamefont
  {Dale}}]{Gould2022-HL}%
  \BibitemOpen
  \bibfield  {author} {\bibinfo {author} {\bibfnamefont {T.}~\bibnamefont
  {Gould}}, \bibinfo {author} {\bibfnamefont {Z.}~\bibnamefont {Hashimi}},
  \bibinfo {author} {\bibfnamefont {L.}~\bibnamefont {Kronik}},\ and\ \bibinfo
  {author} {\bibfnamefont {S.~G.}\ \bibnamefont {Dale}},\ }\bibfield  {title}
  {\bibinfo {title} {Single excitation energies obtained from the ensemble
  {\textquotedblleft}{HOMO}{\textendash}{LUMO} gap{\textquotedblright}: Exact
  results and approximations},\ }\href
  {https://doi.org/10.1021/acs.jpclett.2c00042} {\bibfield  {journal} {\bibinfo
   {journal} {J Phys Chem Lett}\ }\textbf {\bibinfo {volume} {13}},\ \bibinfo
  {pages} {2452} (\bibinfo {year} {2022})}\BibitemShut {NoStop}%
\bibitem [{\citenamefont {Filatov}\ \emph {et~al.}(2015)\citenamefont
  {Filatov}, \citenamefont {Huix-Rotllant},\ and\ \citenamefont
  {Burghardt}}]{Filatov2015-Double}%
  \BibitemOpen
  \bibfield  {author} {\bibinfo {author} {\bibfnamefont {M.}~\bibnamefont
  {Filatov}}, \bibinfo {author} {\bibfnamefont {M.}~\bibnamefont
  {Huix-Rotllant}},\ and\ \bibinfo {author} {\bibfnamefont {I.}~\bibnamefont
  {Burghardt}},\ }\bibfield  {title} {\bibinfo {title} {Ensemble density
  functional theory method correctly describes bond dissociation, excited state
  electron transfer, and double excitations},\ }\href@noop {} {\bibfield
  {journal} {\bibinfo  {journal} {J Chem Phys}\ }\textbf {\bibinfo {volume}
  {142}},\ \bibinfo {pages} {184104} (\bibinfo {year} {2015})}\BibitemShut
  {NoStop}%
\bibitem [{\citenamefont {Sagredo}\ and\ \citenamefont
  {Burke}(2018)}]{Sagredo2018}%
  \BibitemOpen
  \bibfield  {author} {\bibinfo {author} {\bibfnamefont {F.}~\bibnamefont
  {Sagredo}}\ and\ \bibinfo {author} {\bibfnamefont {K.}~\bibnamefont
  {Burke}},\ }\bibfield  {title} {\bibinfo {title} {Accurate double excitations
  from ensemble density functional calculations},\ }\href@noop {} {\bibfield
  {journal} {\bibinfo  {journal} {J Chem Phys}\ }\textbf {\bibinfo {volume}
  {149}},\ \bibinfo {pages} {134103} (\bibinfo {year} {2018})}\BibitemShut
  {NoStop}%
\bibitem [{\citenamefont {Marut}\ \emph {et~al.}(2020)\citenamefont {Marut},
  \citenamefont {Senjean}, \citenamefont {Fromager},\ and\ \citenamefont
  {Loos}}]{Marut2020}%
  \BibitemOpen
  \bibfield  {author} {\bibinfo {author} {\bibfnamefont {C.}~\bibnamefont
  {Marut}}, \bibinfo {author} {\bibfnamefont {B.}~\bibnamefont {Senjean}},
  \bibinfo {author} {\bibfnamefont {E.}~\bibnamefont {Fromager}},\ and\
  \bibinfo {author} {\bibfnamefont {P.-F.}\ \bibnamefont {Loos}},\ }\bibfield
  {title} {\bibinfo {title} {Weight dependence of local exchange-correlation
  functionals in ensemble density-functional theory: Double excitations in
  two-electron systems},\ }\href@noop {} {\bibfield  {journal} {\bibinfo
  {journal} {Faraday Discuss}\ }\textbf {\bibinfo {volume} {224}},\ \bibinfo
  {pages} {402} (\bibinfo {year} {2020})}\BibitemShut {NoStop}%
\bibitem [{\citenamefont {Gould}\ \emph {et~al.}(2021)\citenamefont {Gould},
  \citenamefont {Kronik},\ and\ \citenamefont {Pittalis}}]{Gould2021-DoubleX}%
  \BibitemOpen
  \bibfield  {author} {\bibinfo {author} {\bibfnamefont {T.}~\bibnamefont
  {Gould}}, \bibinfo {author} {\bibfnamefont {L.}~\bibnamefont {Kronik}},\ and\
  \bibinfo {author} {\bibfnamefont {S.}~\bibnamefont {Pittalis}},\ }\bibfield
  {title} {\bibinfo {title} {Double excitations in molecules from ensemble
  density functionals: Theory and approximations},\ }\href
  {https://doi.org/10.1103/physreva.104.022803} {\bibfield  {journal} {\bibinfo
   {journal} {Phys Rev A}\ }\textbf {\bibinfo {volume} {104}},\ \bibinfo
  {pages} {022803} (\bibinfo {year} {2021})}\BibitemShut {NoStop}%
\bibitem [{Note1()}]{Note1}%
  \BibitemOpen
  \bibinfo {note} {This use of prescribed weights excludes the case of finite
  temperature (thermal) ensembles,~\cite {Mermin1965} where weights \protect
  \emph {do depend} on energies (and densities) in a non-trivial
  way}\BibitemShut {NoStop}%
\bibitem [{Note2()}]{Note2}%
  \BibitemOpen
  \bibinfo {note} {We consider only ``well-behaved'' densities here for which
  {$v_s[n]$} exists.}\BibitemShut {Stop}%
\bibitem [{\citenamefont {Gould}\ and\ \citenamefont
  {Pittalis}(2020)}]{Gould2020-SP}%
  \BibitemOpen
  \bibfield  {author} {\bibinfo {author} {\bibfnamefont {T.}~\bibnamefont
  {Gould}}\ and\ \bibinfo {author} {\bibfnamefont {S.}~\bibnamefont
  {Pittalis}},\ }\bibfield  {title} {\bibinfo {title} {Density-driven
  correlations in ensemble density functional theory: Insights from simple
  excitations in atoms},\ }\href@noop {} {\bibfield  {journal} {\bibinfo
  {journal} {Aust J Chem}\ }\textbf {\bibinfo {volume} {73}},\ \bibinfo {pages}
  {714} (\bibinfo {year} {2020})}\BibitemShut {NoStop}%
\bibitem [{Note3()}]{Note3}%
  \BibitemOpen
  \bibinfo {note} {Note, to address degenerate states one must vary degenerate
  manifolds so that degenerate states remain equally weighted.~\cite
  {Gould2020-SP} E.g., addressing the first excited state of Be involves
  setting, $\protect \hat {\Gamma }=\protect \allowbreak (1-w)|1s^22s^2\rangle
  \langle 1s^22s^2|\protect \allowbreak +\protect \genfrac
  {}{}{}1{w}{3}(|1s^22s2p_x\rangle \langle 1s^22s2p_x|\protect \allowbreak
  +|1s^22s2p_y\rangle \langle 1s^22s2p_y|\protect \allowbreak
  +|1s^22s2p_z\rangle \langle 1s^22s2p_z|)$, and varying $w$. Then,
  perturbation theory is well-defined around the ensemble Hx~\cite
  {Gould2017-Limits} ($\lambda \to 0^+$) limit.}\BibitemShut {Stop}%
\bibitem [{\citenamefont {Wigner}(1934)}]{Wig-PR-34}%
  \BibitemOpen
  \bibfield  {author} {\bibinfo {author} {\bibfnamefont {E.~P.}\ \bibnamefont
  {Wigner}},\ }\href@noop {} {\bibfield  {journal} {\bibinfo  {journal} {Phys
  Rev}\ }\textbf {\bibinfo {volume} {{46}}},\ \bibinfo {pages} {1002} (\bibinfo
  {year} {1934})}\BibitemShut {NoStop}%
\bibitem [{\citenamefont {Wigner}(1938)}]{Wig-TFS-38}%
  \BibitemOpen
  \bibfield  {author} {\bibinfo {author} {\bibfnamefont {E.~P.}\ \bibnamefont
  {Wigner}},\ }\href@noop {} {\bibfield  {journal} {\bibinfo  {journal} {Trans
  Faraday Soc}\ }\textbf {\bibinfo {volume} {{34}}},\ \bibinfo {pages} {678}
  (\bibinfo {year} {1938})}\BibitemShut {NoStop}%
\bibitem [{Note4()}]{Note4}%
  \BibitemOpen
  \bibinfo {note} {Note, we do not scale the mixing weights. Also note that,
  level crossings in the ensemble are not a concern as excitation energies do
  not change order under uniform scaling.}\BibitemShut {Stop}%
\bibitem [{Note5()}]{Note5}%
  \BibitemOpen
  \bibinfo {note} {This result is easily shown by taking $\gamma ^2$ times a
  series expansion of eq.~\protect \textup {\hbox {\mathsurround \z@ \protect
  \normalfont (\ignorespaces \ref {eqn:ACF}\unskip \@@italiccorr )}} in small
  $\lambda =\gamma ^{-1}$.}\BibitemShut {Stop}%
\bibitem [{\citenamefont {Görling}\ and\ \citenamefont
  {Levy}(1993)}]{Goerling1993}%
  \BibitemOpen
  \bibfield  {author} {\bibinfo {author} {\bibfnamefont {A.}~\bibnamefont
  {Görling}}\ and\ \bibinfo {author} {\bibfnamefont {M.}~\bibnamefont
  {Levy}},\ }\bibfield  {title} {\bibinfo {title} {Correlation-energy
  functional and its high-density limit obtained from a coupling-constant
  perturbation expansion},\ }\href {https://doi.org/10.1103/physrevb.47.13105}
  {\bibfield  {journal} {\bibinfo  {journal} {Phys Rev B}\ }\textbf {\bibinfo
  {volume} {47}},\ \bibinfo {pages} {13105} (\bibinfo {year}
  {1993})}\BibitemShut {NoStop}%
\bibitem [{\citenamefont {Yang}(2021)}]{Yang2021}%
  \BibitemOpen
  \bibfield  {author} {\bibinfo {author} {\bibfnamefont {Z.}~\bibnamefont
  {Yang}},\ }\bibfield  {title} {\bibinfo {title} {Second-order perturbative
  correlation energy functional in the ensemble density-functional theory},\
  }\href {https://doi.org/10.1103/physreva.104.052806} {\bibfield  {journal}
  {\bibinfo  {journal} {Phys Rev A}\ }\textbf {\bibinfo {volume} {104}},\
  \bibinfo {pages} {052806} (\bibinfo {year} {2021})}\BibitemShut {NoStop}%
\bibitem [{\citenamefont {Gould}\ and\ \citenamefont
  {Pittalis}(2019)}]{Gould2019-DD}%
  \BibitemOpen
  \bibfield  {author} {\bibinfo {author} {\bibfnamefont {T.}~\bibnamefont
  {Gould}}\ and\ \bibinfo {author} {\bibfnamefont {S.}~\bibnamefont
  {Pittalis}},\ }\bibfield  {title} {\bibinfo {title} {Density-driven
  correlations in many-electron ensembles: Theory and application for excited
  states},\ }\href@noop {} {\bibfield  {journal} {\bibinfo  {journal} {Phys Rev
  Lett}\ }\textbf {\bibinfo {volume} {123}},\ \bibinfo {pages} {016401}
  (\bibinfo {year} {2019})}\BibitemShut {NoStop}%
\bibitem [{\citenamefont {Fromager}(2020)}]{Fromager2020-DD}%
  \BibitemOpen
  \bibfield  {author} {\bibinfo {author} {\bibfnamefont {E.}~\bibnamefont
  {Fromager}},\ }\bibfield  {title} {\bibinfo {title} {Individual correlations
  in ensemble density-functional theory: State-driven/density-driven
  decompositions without additional {Kohn-Sham} systems},\ }\href@noop {}
  {\bibfield  {journal} {\bibinfo  {journal} {Phys Rev Lett}\ }\textbf
  {\bibinfo {volume} {124}},\ \bibinfo {pages} {243001} (\bibinfo {year}
  {2020})}\BibitemShut {NoStop}%
\bibitem [{Note6()}]{Note6}%
  \BibitemOpen
  \bibinfo {note} {This is $V_{\protect \rm cl}=\protect \genfrac
  {}{}{}1{\lambda }{2R_0}$ where $R_0$ minimizes the classical energy,
  $E_{\protect \rm cl}(R)=2\times \protect \frac {1}{2}R^2 + \protect \genfrac
  {}{}{}1{\lambda }{2R}$, of two electrons interacting with $\protect \genfrac
  {}{}{}1{\lambda }{|\protect \boldsymbol {R}_1-\protect \boldsymbol {R}_2|}$
  when the two electrons are on opposite sides of the well.}\BibitemShut
  {Stop}%
\bibitem [{\citenamefont {Colombo}\ \emph {et~al.}(2021)\citenamefont
  {Colombo}, \citenamefont {Di~Marino},\ and\ \citenamefont
  {Stra}}]{ColDMaStra-arxiv-21}%
  \BibitemOpen
  \bibfield  {author} {\bibinfo {author} {\bibfnamefont {M.}~\bibnamefont
  {Colombo}}, \bibinfo {author} {\bibfnamefont {S.}~\bibnamefont {Di~Marino}},\
  and\ \bibinfo {author} {\bibfnamefont {F.}~\bibnamefont {Stra}},\ }\bibfield
  {title} {\bibinfo {title} {First order expansion in the semiclassical limit
  of the levy-lieb functional},\ }\href@noop {} {\bibfield  {journal} {\bibinfo
   {journal} {arXiv preprint arXiv:2106.06282}\ } (\bibinfo {year}
  {2021})}\BibitemShut {NoStop}%
\bibitem [{\citenamefont {Grossi}\ \emph {et~al.}(2019)\citenamefont {Grossi},
  \citenamefont {Seidl}, \citenamefont {Gori-Giorgi},\ and\ \citenamefont
  {Giesbertz}}]{Grossi2019}%
  \BibitemOpen
  \bibfield  {author} {\bibinfo {author} {\bibfnamefont {J.}~\bibnamefont
  {Grossi}}, \bibinfo {author} {\bibfnamefont {M.}~\bibnamefont {Seidl}},
  \bibinfo {author} {\bibfnamefont {P.}~\bibnamefont {Gori-Giorgi}},\ and\
  \bibinfo {author} {\bibfnamefont {K.~J.~H.}\ \bibnamefont {Giesbertz}},\
  }\bibfield  {title} {\bibinfo {title} {Functional derivative of the
  zero-point-energy functional from the strong-interaction limit of
  density-functional theory},\ }\href
  {https://doi.org/10.1103/physreva.99.052504} {\bibfield  {journal} {\bibinfo
  {journal} {Phys Rev A}\ }\textbf {\bibinfo {volume} {99}},\ \bibinfo {pages}
  {052504} (\bibinfo {year} {2019})}\BibitemShut {NoStop}%
\bibitem [{\citenamefont {Harriman}(1981)}]{Harriman1981}%
  \BibitemOpen
  \bibfield  {author} {\bibinfo {author} {\bibfnamefont {J.~E.}\ \bibnamefont
  {Harriman}},\ }\bibfield  {title} {\bibinfo {title} {Orthonormal orbitals for
  the representation of an arbitrary density},\ }\href
  {https://doi.org/10.1103/physreva.24.680} {\bibfield  {journal} {\bibinfo
  {journal} {Phys Rev A}\ }\textbf {\bibinfo {volume} {24}},\ \bibinfo {pages}
  {680} (\bibinfo {year} {1981})}\BibitemShut {NoStop}%
\bibitem [{\citenamefont {Savin}(1995)}]{Sav-PRA-95}%
  \BibitemOpen
  \bibfield  {author} {\bibinfo {author} {\bibfnamefont {A.}~\bibnamefont
  {Savin}},\ }\bibfield  {title} {\bibinfo {title} {Expression of the exact
  electron-correlation-energy density functional in terms of first-order
  density matrices},\ }\href {https://doi.org/10.1103/PhysRevA.52.R1805}
  {\bibfield  {journal} {\bibinfo  {journal} {Phys Rev A}\ }\textbf {\bibinfo
  {volume} {52}},\ \bibinfo {pages} {R1805} (\bibinfo {year}
  {1995})}\BibitemShut {NoStop}%
\bibitem [{\citenamefont {Levy}\ and\ \citenamefont
  {G\"orling}(1995)}]{LevGor-PRA-95}%
  \BibitemOpen
  \bibfield  {author} {\bibinfo {author} {\bibfnamefont {M.}~\bibnamefont
  {Levy}}\ and\ \bibinfo {author} {\bibfnamefont {A.}~\bibnamefont
  {G\"orling}},\ }\bibfield  {title} {\bibinfo {title} {Correlation-energy
  density-functional formulas from correlating first-order density matrices},\
  }\href {https://doi.org/10.1103/PhysRevA.52.R1808} {\bibfield  {journal}
  {\bibinfo  {journal} {Phys Rev A}\ }\textbf {\bibinfo {volume} {52}},\
  \bibinfo {pages} {R1808} (\bibinfo {year} {1995})}\BibitemShut {NoStop}%
\bibitem [{\citenamefont {Teale}\ \emph {et~al.}(2016)\citenamefont {Teale},
  \citenamefont {Helgaker},\ and\ \citenamefont {Savin}}]{TeaHelSav-JCCS-16}%
  \BibitemOpen
  \bibfield  {author} {\bibinfo {author} {\bibfnamefont {A.~M.}\ \bibnamefont
  {Teale}}, \bibinfo {author} {\bibfnamefont {T.}~\bibnamefont {Helgaker}},\
  and\ \bibinfo {author} {\bibfnamefont {A.}~\bibnamefont {Savin}},\ }\bibfield
   {title} {\bibinfo {title} {Alternative representations of the correlation
  energy in density-functional theory: A kinetic-energy based adiabatic
  connection},\ }\href {https://doi.org/10.1002/jccs.201500132} {\bibfield
  {journal} {\bibinfo  {journal} {J. Chin. Chem. Soc.}\ }\textbf {\bibinfo
  {volume} {63}},\ \bibinfo {pages} {121} (\bibinfo {year} {2016})}\BibitemShut
  {NoStop}%
\bibitem [{\citenamefont {Gori-Giorgi}\ and\ \citenamefont
  {Seidl}(2010)}]{GorSei-PCCP-10}%
  \BibitemOpen
  \bibfield  {author} {\bibinfo {author} {\bibfnamefont {P.}~\bibnamefont
  {Gori-Giorgi}}\ and\ \bibinfo {author} {\bibfnamefont {M.}~\bibnamefont
  {Seidl}},\ }\bibfield  {title} {\bibinfo {title} {Density functional theory
  for strongly-interacting electrons: Perspectives for physics and chemistry},\
  }\href {https://doi.org/10.1039/c0cp01061h} {\bibfield  {journal} {\bibinfo
  {journal} {Phys. Chem. Chem. Phys}\ }\textbf {\bibinfo {volume} {12}},\
  \bibinfo {pages} {14405} (\bibinfo {year} {2010})}\BibitemShut {NoStop}%
\bibitem [{\citenamefont {Deur}\ \emph {et~al.}(2018)\citenamefont {Deur},
  \citenamefont {Mazouin}, \citenamefont {Senjean},\ and\ \citenamefont
  {Fromager}}]{Deur2018}%
  \BibitemOpen
  \bibfield  {author} {\bibinfo {author} {\bibfnamefont {K.}~\bibnamefont
  {Deur}}, \bibinfo {author} {\bibfnamefont {L.}~\bibnamefont {Mazouin}},
  \bibinfo {author} {\bibfnamefont {B.}~\bibnamefont {Senjean}},\ and\ \bibinfo
  {author} {\bibfnamefont {E.}~\bibnamefont {Fromager}},\ }\bibfield  {title}
  {\bibinfo {title} {Exploring weight-dependent density-functional
  approximations for ensembles in the hubbard dimer},\ }\bibfield  {journal}
  {\bibinfo  {journal} {Eur. Phys. J. B}\ }\textbf {\bibinfo {volume} {91}},\
  \href {https://doi.org/10.1140/epjb/e2018-90124-7}
  {10.1140/epjb/e2018-90124-7} (\bibinfo {year} {2018})\BibitemShut {NoStop}%
\bibitem [{\citenamefont {Maitra}(2022)}]{Maitra2022-AR}%
  \BibitemOpen
  \bibfield  {author} {\bibinfo {author} {\bibfnamefont {N.~T.}\ \bibnamefont
  {Maitra}},\ }\bibfield  {title} {\bibinfo {title} {Double and charge-transfer
  excitations in time-dependent density functional theory},\ }\href
  {https://doi.org/10.1146/annurev-physchem-082720-124933} {\bibfield
  {journal} {\bibinfo  {journal} {Annu. Rev. Phys. Chem.}\ }\textbf {\bibinfo
  {volume} {73}},\ \bibinfo {pages} {117} (\bibinfo {year} {2022})},\ \bibinfo
  {note} {pMID: 34910562}\BibitemShut {NoStop}%
\bibitem [{\citenamefont {Giesbertz}\ \emph {et~al.}(2008)\citenamefont
  {Giesbertz}, \citenamefont {Baerends},\ and\ \citenamefont
  {Gritsenko}}]{Giesbertz2008-CTDBB}%
  \BibitemOpen
  \bibfield  {author} {\bibinfo {author} {\bibfnamefont {K.~J.~H.}\
  \bibnamefont {Giesbertz}}, \bibinfo {author} {\bibfnamefont {E.~J.}\
  \bibnamefont {Baerends}},\ and\ \bibinfo {author} {\bibfnamefont {O.~V.}\
  \bibnamefont {Gritsenko}},\ }\bibfield  {title} {\bibinfo {title} {Charge
  transfer, double and bond-breaking excitations with time-dependent density
  matrix functional theory},\ }\href
  {https://doi.org/10.1103/PhysRevLett.101.033004} {\bibfield  {journal}
  {\bibinfo  {journal} {Phys. Rev. Lett.}\ }\textbf {\bibinfo {volume} {101}},\
  \bibinfo {pages} {033004} (\bibinfo {year} {2008})}\BibitemShut {NoStop}%
\bibitem [{\citenamefont {Seidl}\ \emph {et~al.}(2000)\citenamefont {Seidl},
  \citenamefont {Perdew},\ and\ \citenamefont {Kurth}}]{Seidl2000}%
  \BibitemOpen
  \bibfield  {author} {\bibinfo {author} {\bibfnamefont {M.}~\bibnamefont
  {Seidl}}, \bibinfo {author} {\bibfnamefont {J.~P.}\ \bibnamefont {Perdew}},\
  and\ \bibinfo {author} {\bibfnamefont {S.}~\bibnamefont {Kurth}},\ }\bibfield
   {title} {\bibinfo {title} {Density functionals for the strong-interaction
  limit},\ }\href {https://doi.org/10.1103/physreva.62.012502} {\bibfield
  {journal} {\bibinfo  {journal} {Phys Rev A}\ }\textbf {\bibinfo {volume}
  {62}},\ \bibinfo {pages} {012502} (\bibinfo {year} {2000})}\BibitemShut
  {NoStop}%
\bibitem [{\citenamefont {{\'{S}}miga}\ \emph {et~al.}(2022)\citenamefont
  {{\'{S}}miga}, \citenamefont {Sala}, \citenamefont {Gori-Giorgi},\ and\
  \citenamefont {Fabiano}}]{SDGF22}%
  \BibitemOpen
  \bibfield  {author} {\bibinfo {author} {\bibfnamefont {S.}~\bibnamefont
  {{\'{S}}miga}}, \bibinfo {author} {\bibfnamefont {F.~D.}\ \bibnamefont
  {Sala}}, \bibinfo {author} {\bibfnamefont {P.}~\bibnamefont {Gori-Giorgi}},\
  and\ \bibinfo {author} {\bibfnamefont {E.}~\bibnamefont {Fabiano}},\
  }\bibfield  {title} {\bibinfo {title} {Self-consistent implementation of
  {Kohn-Sham} adiabatic connection models with improved treatment of the
  strong-interaction limit},\ }\bibfield  {journal} {\bibinfo  {journal} {J
  Chem Theory Comput}\ }\href {https://doi.org/10.1021/acs.jctc.2c00352}
  {10.1021/acs.jctc.2c00352} (\bibinfo {year} {2022})\BibitemShut {NoStop}%
\bibitem [{\citenamefont {Marie}\ \emph {et~al.}(2022)\citenamefont {Marie},
  \citenamefont {Kooi}, \citenamefont {Grossi}, \citenamefont {Seidl},
  \citenamefont {Musslimani}, \citenamefont {Giesbertz},\ and\ \citenamefont
  {Gori-Giorgi}}]{Marie2022}%
  \BibitemOpen
  \bibfield  {author} {\bibinfo {author} {\bibfnamefont {A.}~\bibnamefont
  {Marie}}, \bibinfo {author} {\bibfnamefont {D.~P.}\ \bibnamefont {Kooi}},
  \bibinfo {author} {\bibfnamefont {J.}~\bibnamefont {Grossi}}, \bibinfo
  {author} {\bibfnamefont {M.}~\bibnamefont {Seidl}}, \bibinfo {author}
  {\bibfnamefont {Z.~H.}\ \bibnamefont {Musslimani}}, \bibinfo {author}
  {\bibfnamefont {K.}~\bibnamefont {Giesbertz}},\ and\ \bibinfo {author}
  {\bibfnamefont {P.}~\bibnamefont {Gori-Giorgi}},\ }\href
  {https://doi.org/10.48550/ARXIV.2208.14546} {\bibinfo {title} {Real space
  mott-anderson electron localization with long-range interactions: exact and
  approximate descriptions}} (\bibinfo {year} {2022})\BibitemShut {NoStop}%
\bibitem [{\citenamefont {Kirkpatrick}\ \emph {et~al.}(2021)\citenamefont
  {Kirkpatrick}, \citenamefont {McMorrow}, \citenamefont {Turban},
  \citenamefont {Gaunt}, \citenamefont {Spencer}, \citenamefont {Matthews},
  \citenamefont {Obika}, \citenamefont {Thiry}, \citenamefont {Fortunato},
  \citenamefont {Pfau}, \citenamefont {Castellanos}, \citenamefont {Petersen},
  \citenamefont {Nelson}, \citenamefont {Kohli}, \citenamefont
  {Mori-S{\'{a}}nchez}, \citenamefont {Hassabis},\ and\ \citenamefont
  {Cohen}}]{DM21}%
  \BibitemOpen
  \bibfield  {author} {\bibinfo {author} {\bibfnamefont {J.}~\bibnamefont
  {Kirkpatrick}}, \bibinfo {author} {\bibfnamefont {B.}~\bibnamefont
  {McMorrow}}, \bibinfo {author} {\bibfnamefont {D.~H.~P.}\ \bibnamefont
  {Turban}}, \bibinfo {author} {\bibfnamefont {A.~L.}\ \bibnamefont {Gaunt}},
  \bibinfo {author} {\bibfnamefont {J.~S.}\ \bibnamefont {Spencer}}, \bibinfo
  {author} {\bibfnamefont {A.~G. D.~G.}\ \bibnamefont {Matthews}}, \bibinfo
  {author} {\bibfnamefont {A.}~\bibnamefont {Obika}}, \bibinfo {author}
  {\bibfnamefont {L.}~\bibnamefont {Thiry}}, \bibinfo {author} {\bibfnamefont
  {M.}~\bibnamefont {Fortunato}}, \bibinfo {author} {\bibfnamefont
  {D.}~\bibnamefont {Pfau}}, \bibinfo {author} {\bibfnamefont {L.~R.}\
  \bibnamefont {Castellanos}}, \bibinfo {author} {\bibfnamefont
  {S.}~\bibnamefont {Petersen}}, \bibinfo {author} {\bibfnamefont {A.~W.~R.}\
  \bibnamefont {Nelson}}, \bibinfo {author} {\bibfnamefont {P.}~\bibnamefont
  {Kohli}}, \bibinfo {author} {\bibfnamefont {P.}~\bibnamefont
  {Mori-S{\'{a}}nchez}}, \bibinfo {author} {\bibfnamefont {D.}~\bibnamefont
  {Hassabis}},\ and\ \bibinfo {author} {\bibfnamefont {A.~J.}\ \bibnamefont
  {Cohen}},\ }\bibfield  {title} {\bibinfo {title} {Pushing the frontiers of
  density functionals by solving the fractional electron problem},\ }\href
  {https://doi.org/10.1126/science.abj6511} {\bibfield  {journal} {\bibinfo
  {journal} {Science}\ }\textbf {\bibinfo {volume} {374}},\ \bibinfo {pages}
  {1385} (\bibinfo {year} {2021})}\BibitemShut {NoStop}%
\bibitem [{\citenamefont {Mermin}(1965)}]{Mermin1965}%
  \BibitemOpen
  \bibfield  {author} {\bibinfo {author} {\bibfnamefont {N.~D.}\ \bibnamefont
  {Mermin}},\ }\bibfield  {title} {\bibinfo {title} {Thermal properties of the
  inhomogeneous electron gas},\ }\href
  {https://doi.org/10.1103/physrev.137.a1441} {\bibfield  {journal} {\bibinfo
  {journal} {Phys Rev}\ }\textbf {\bibinfo {volume} {137}},\ \bibinfo {pages}
  {A1441} (\bibinfo {year} {1965})}\BibitemShut {NoStop}%
\bibitem [{\citenamefont {Pittalis}\ \emph {et~al.}(2011)\citenamefont
  {Pittalis}, \citenamefont {Proetto}, \citenamefont {Floris}, \citenamefont
  {Sanna}, \citenamefont {Bersier}, \citenamefont {Burke},\ and\ \citenamefont
  {Gross}}]{Pittalis2011-FT}%
  \BibitemOpen
  \bibfield  {author} {\bibinfo {author} {\bibfnamefont {S.}~\bibnamefont
  {Pittalis}}, \bibinfo {author} {\bibfnamefont {C.~R.}\ \bibnamefont
  {Proetto}}, \bibinfo {author} {\bibfnamefont {A.}~\bibnamefont {Floris}},
  \bibinfo {author} {\bibfnamefont {A.}~\bibnamefont {Sanna}}, \bibinfo
  {author} {\bibfnamefont {C.}~\bibnamefont {Bersier}}, \bibinfo {author}
  {\bibfnamefont {K.}~\bibnamefont {Burke}},\ and\ \bibinfo {author}
  {\bibfnamefont {E.~K.~U.}\ \bibnamefont {Gross}},\ }\bibfield  {title}
  {\bibinfo {title} {Exact conditions in finite-temperature density-functional
  theory},\ }\href {https://doi.org/10.1103/PhysRevLett.107.163001} {\bibfield
  {journal} {\bibinfo  {journal} {Phys. Rev. Lett.}\ }\textbf {\bibinfo
  {volume} {107}},\ \bibinfo {pages} {163001} (\bibinfo {year}
  {2011})}\BibitemShut {NoStop}%
\bibitem [{\citenamefont {Vignale}\ and\ \citenamefont
  {Rasolt}(1988)}]{Vignale1988}%
  \BibitemOpen
  \bibfield  {author} {\bibinfo {author} {\bibfnamefont {G.}~\bibnamefont
  {Vignale}}\ and\ \bibinfo {author} {\bibfnamefont {M.}~\bibnamefont
  {Rasolt}},\ }\bibfield  {title} {\bibinfo {title} {Current- and
  spin-density-functional theory for inhomogeneous electronic systems in strong
  magnetic fields},\ }\href {https://doi.org/10.1103/PhysRevB.37.10685}
  {\bibfield  {journal} {\bibinfo  {journal} {Phys. Rev. B}\ }\textbf {\bibinfo
  {volume} {37}},\ \bibinfo {pages} {10685} (\bibinfo {year}
  {1988})}\BibitemShut {NoStop}%
\bibitem [{\citenamefont {Bencheikh}(2003)}]{Bencheikh_2003}%
  \BibitemOpen
  \bibfield  {author} {\bibinfo {author} {\bibfnamefont {K.}~\bibnamefont
  {Bencheikh}},\ }\bibfield  {title} {\bibinfo {title} {Spin{\textendash}orbit
  coupling in the spin-current-density-functional theory},\ }\href
  {https://doi.org/10.1088/0305-4470/36/48/002} {\bibfield  {journal} {\bibinfo
   {journal} {J. Phys. A: Math. Gen.}\ }\textbf {\bibinfo {volume} {36}},\
  \bibinfo {pages} {11929} (\bibinfo {year} {2003})}\BibitemShut {NoStop}%
\bibitem [{\citenamefont {Pittalis}\ \emph {et~al.}(2017)\citenamefont
  {Pittalis}, \citenamefont {Vignale},\ and\ \citenamefont
  {Eich}}]{Pittalis-U1SU2}%
  \BibitemOpen
  \bibfield  {author} {\bibinfo {author} {\bibfnamefont {S.}~\bibnamefont
  {Pittalis}}, \bibinfo {author} {\bibfnamefont {G.}~\bibnamefont {Vignale}},\
  and\ \bibinfo {author} {\bibfnamefont {F.~G.}\ \bibnamefont {Eich}},\
  }\bibfield  {title} {\bibinfo {title}
  {$\text{U}(1)\ifmmode\times\else\texttimes\fi{}\mathrm{SU}(2)$ gauge
  invariance made simple for density functional approximations},\ }\href
  {https://doi.org/10.1103/PhysRevB.96.035141} {\bibfield  {journal} {\bibinfo
  {journal} {Phys. Rev. B}\ }\textbf {\bibinfo {volume} {96}},\ \bibinfo
  {pages} {035141} (\bibinfo {year} {2017})}\BibitemShut {NoStop}%
\end{thebibliography}
